\documentclass[aps,prl,twocolumn,superscriptaddress,showpacs]{revtex4}

\usepackage{graphicx}
\usepackage{color}
\usepackage{bm}

\DeclareTextSymbol{\degre}{T1}{6}
\DeclareTextSymbol{\degre}{OT1}{23}
\DeclareGraphicsExtensions{.jpg,.eps}
\newcommand{\QSR}{$q=\sqrt{3}\times\sqrt{3}$}
\newcommand{\QO}{$q=0$}
\newcommand{\ud}{\mathrm{d}}

\begin{document}

\title{Propagation and Ghosts in the Classical Kagome Antiferromagnet}
\author{J. Robert}
\affiliation{Laboratoire L\'eon Brillouin, CEA-CNRS, CE-Saclay, 91191
Gif-sur-Yvette, France}
\author{B. Canals}
\author{V. Simonet}
\author{R. Ballou}
\affiliation{Institut NEEL, CNRS \& Universit\'e Joseph Fourier, BP
166, 38042 Grenoble Cedex 9, France}

\date{\today}

\begin{abstract}
We investigate the classical spin dynamics of the kagome
antiferromagnet by combining Monte Carlo and spin dynamics
simulations. We show that this model has two distinct low
temperature dynamical regimes, both sustaining propagative modes.
The expected gauge invariance type of the low energy low
temperature out of plane excitations is also evidenced in the non
linear regime. A detailed analysis of the excitations allows to
identify ghosts in the dynamical structure factor, i.e propagating
excitations with a strongly reduced spectral weight. We argue that
these dynamical extinction rules are of geometrical origin.
\end{abstract}


\maketitle


Geometrical frustrated magnets are currently a source of high
interest for the exotic phases and unexpected dynamics that they
are liable to generate. A full insight about their behaviors is
still far from having been acquired, in particular at the lowest
temperatures.

A prototype is the classical Heisenberg kagome
antiferromagnet \cite{kagome}. As a basic distinctive feature of
the geometrical frustration, its ground state consists in a
continuous connected manifold of spin configurations. At high
temperatures ($T/J \gtrsim 0.1$, with $J$ the first neighbor
exchange), the system is paramagnetic. It enters what we shall
call from now on a cooperative magnetic phase in the range $5\cdot
10^{-3} \lesssim T/J \lesssim 0.1$ where short range correlations
are enhanced. At the lowest temperatures ($T/J \lesssim 5\cdot
10^{-3}$), thermal fluctuations above each of the spin
configurations of the ground state manifold are not equivalent and
drive an entropic based order out of disorder mechanism
\cite{villain1979}, eventually selecting a spin
plane \cite{chalker1992} and developing an octupolar order
\cite{zhitomirsky2008}. We shall call this phase coplanar to
distinguish it from the former. While in both low temperature
regimes it was shown that spin pair correlations remain short
ranged \cite{reimers1993}, it is only in the coplanar  phase  that
the continuous degeneracy of the manifold was argued to be reduced
to a discrete one, described by the 3-colorings of the lattice
\cite{huse1992}. Altogether, these results provide a rather clear
picture of the thermodynamics of the classical kagome
antiferromagnet, which should apply to experimental compounds with
large magnetic moments but also be of some relevance for low spin
systems, since quantum fluctuations often play a significant role
at very low temperatures only.

A much poorer understanding of the spin dynamics is in contrast
available. To our knowledge, only one numerical investigation was
so far conducted \cite{keren1994}, which furthermore was not
resolved in momentum vectors ${\mathbf Q}$, thus ignoring any diffusive
or propagating aspects of the excitations. In this letter, we
analyze the temperature dependent dynamics of the classical kagome
antiferromagnet from two point of views. We first show that at low
temperatures, spin waves (SW) do propagate and are sensitive to
the underlying spin texture, either cooperative paramagnetic or
coplanar. Quantitative analysis of the dynamical structure factor
is performed and provides the characteristic time scales.
Additionally, the invariance of the linear SW spectra with respect
to the ground state spin configurations on which they are built is
evidenced in a wide range of temperatures, including those where
non linear effects are at play. We next put forward that peculiar
excitations develop that would be almost invisible to dynamical
spin-pair correlations sensitive probes, such as inelastic neutron
scattering.

The numerical method used in this work is a combination of an
hybrid Monte Carlo (MC) method, which allows generating samples of
spin arrays at a given temperature, and an integration of the
non-linear coupled equations of motion for the spin dynamics (SD):
\begin{equation}
\frac{\ud \mathbf{S}_i}{\ud t}=J\,  \bigg( \sum_{j}\mathbf{S}_j
\bigg) \times \mathbf{S}_i, \label{eq:eq_mvt}
\end{equation}
where $j$ is a first neighbor of $i$ and $J>0$ is the
antiferromagnetic exchange \cite{tsai2000}. The numerical
integration has been performed up to $t=1000\,J^{-1}$ using an
8$^{\mathrm{th}}$-order Runge-Kutta method (RK) with an adaptative
step-size control. The RK error parameter as well as the RK order
have been fixed in order to preserve the euclidian distance with a
test-full diffusion of Eq. \ref{eq:eq_mvt} performed with the more
robust but time consuming Burlisch-Stoer algorithm. As a result,
trivial constants of motion, such as the total energy $E_{tot}$
and magnetization $M_{tot}$, are conserved with a relative error
smaller than $10^{-6}$. As for the spin arrays samplings by the MC
method, a first run has been performed in order to find an optimal
set of temperatures for a parallel tempering scheme
\cite{hukushima1996}, which minimizes the ergodic time
\cite{katzgraber2006}. A reduction of the solid angle for each
spin flip trial together with rotations around the local molecular
fields ensure a rate acceptance above 40\%. The numerical
simulations reported in this work were performed on samples of $L
\times L \times 3$ spins with $L = 36$ and periodic boundary
conditions. Our interest lies in the scattering function, namely
the time and space Fourier transform of the dynamical spin-pair
correlations:
\begin{eqnarray}
S(\mathbf{Q},\omega) = \sum_{ij} \int \frac{\ud t}{\sqrt{2\pi}N}
\, \langle \mathbf{S}_i(0) \cdot \mathbf{S}_j(t) \rangle \,
\mathrm{e}^{-i\mathbf{Q}\cdot \mathbf{R}_{ij}} \,
\mathrm{e}^{-i\omega t}
\end{eqnarray}
where $\mathbf{Q}$ and $\omega$ are the momentum vector and energy
transfer, $\langle ... \rangle$ is the ensemble average,
$\mathbf{R}_{ij}=\mathbf{R}_{j}-\mathbf{R}_{i}$ and $N$ the number
of spins.

\begin{figure}[t!]
\includegraphics[width=8.66cm]{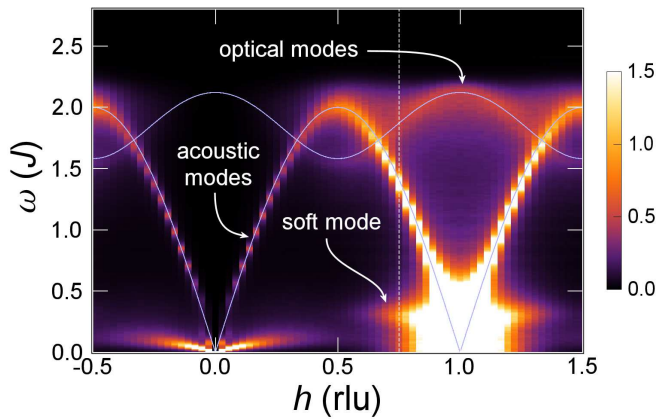}
\caption{(color online) Intensity map $(a.u.)$ of the scattering
function \textit{vs.} $\omega$ and $\mathbf{Q} = (h,0)$, for $T/J=5\cdot
10^{-4}$. The LSW dispersion relations $\omega(\mathbf{Q})$ for
the phase \QSR~ are plotted as blue lines. The white dotted
line corresponds to the constant-$\mathbf{Q}$ scans presented on
Fig. \ref{fig:fig2}.} \label{fig:fig1}
\end{figure}


Details about static properties will not be given here. It however
is worth noting that our results for $\omega = 0$ at very low
temperatures ($T/J \lesssim 5\cdot 10^{-3}$), when entropic
selection is at work, points towards fluctuations predominantly
associated with the so-called \QSR\ phase. This meets with
previous conclusions, although those were inferred from
instantaneous ensemble averages
\cite{harris1992,huse1992,reimers1993}. We similarly got very good
agreements with previous numerical or analytical investigations of
the specific heat, the coplanar ordering, or the instantaneous
scattering function
\cite{reimers1993,chalker1992,zhitomirsky2008,garanin1999}, giving
confidence on the quality of our numerical simulations.

Let us now focus on the dynamical properties of the kagome
antiferromagnet. Although the propagation of collective
excitations may appear unexpected in such a system, where the
spin-pair correlation function decays exponentially with distance
at finite temperatures \cite{reimers1993}, a sufficient temporal
and spatial stiffness may lead to the propagation of SW in locally
ordered regions. Therefore, a required condition for the
development of SW excitations is an increase with decreasing
temperature of the autocorrelation time $\tau_a$ featuring the
lifetime of locally ordered states. $\tau_a$ has been numerically
evaluated by integrating the scattering function over all
$\mathbf{Q}$-values in the reciprocal space, which gives access to
the time Fourier transform of the autocorrelation function
$A(t)=\langle \mathbf{S}_i(0) \cdot \mathbf{S}_i(t)\rangle$. A fit
of the obtained quasielastic (QE) signal using a lorentzian shape
$\frac{I_0\Gamma_a}{\Gamma_a^2+\omega^2}$, associated with a
decaying exponential law $A(t)=\exp{(-\Gamma_a t)}$ in time space,
allows extracting the Half Width at Half Maximum (HWHM) $\Gamma_a
\propto 1/\tau_a$. As the temperature is decreased, this shows an
algebraic variation $\Gamma_a=\mathcal{A}T^{\zeta}$ with $\zeta =
0.995 \pm 0.018$ for $T/J \lesssim 0.1$ (see Fig. \ref{fig:fig3}a), transposing to a slowing down of the spin fluctuations, but
nevertheless no spin freezing even at temperatures as low as
$T/J=5\cdot 10^{-4}$. Interestingly, the same thermal
variation is observed in the cooperative paramagnetic ($T/J\gtrsim
5\cdot 10^{-3}$) and the coplanar states ($T/J\lesssim 5\cdot
10^{-3}$) regimes, asserting that the entropic
selection favoring the coplanar manifold has no influence on the
lifetime $\tau_a$ of locally ordered states. Now that we have
characterized the temporal stiffness associated with the $T^{-1}$
slowing down of $\tau_a$, we address the question of well defined
excitations as well as their possible propagation. An evidence of
the existence of SW-type excitations at low temperatures is
explicit in the excitation spectrum for $T/J=5\cdot 10^{-4}$ (see
Fig. \ref{fig:fig1}). For comparison, the linear spin wave (LSW)
spectrum \cite{harris1992}  emerging from the pure \QSR\ phase is
shown. The SD simulations evidence a large weight of
$S(\mathbf{Q},\omega)$ at this LSW spectrum, confirming that the
\QSR\ short range dynamical correlations are favored at very low
temperature.

\begin{figure}[b!]
\includegraphics[bb=41 19 300 231, scale=0.9]{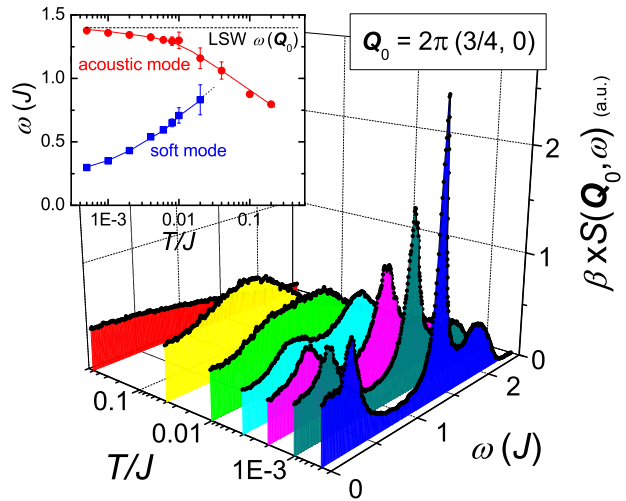}
\caption{(color online) Temperature weighted scattering function
\textit{vs.} energy at different temperatures for
$\mathbf{Q}_0=2\pi(3/4,0)$. Inset : position of soft and acoustic
modes versus temperature; error bars obtained from
several fitting processes.} \label{fig:fig2}
\end{figure}

The analysis of the spectrum as a function of the temperature
allows getting more insights about the formation of the SW
excitations. We show in Fig. \ref{fig:fig2} the frequency
dependence of $S(\mathbf{Q},\omega)$ at the point
$\mathbf{Q}_0=2\pi(3/4,0)$ in the  reciprocal space, located
between the Brillouin zone (BZ) boundary and the BZ center
(see Fig. \ref{fig:fig1}), where the soft, acoustic and optical
modes are particularly easy to distinguish. These
constant-$\mathbf{Q}$ scans are represented for temperatures from
$T/J=0.5$ to $5\cdot 10^{-4}$. At high temperatures ($T/J\gtrsim
0.2$), only a QE signal centered at $\omega=0$ contributes to the
scattering function $S(\mathbf{Q},\omega)$. A single broad
excitation at finite energy comes into sight on decreasing $T/J$
from $0.2$ to $10^{-2}$, although strongly softened compared to
the LSW theory expectation (see inset of Fig. \ref{fig:fig2}).
Below $T/J=10^{-2}$, the broad peak splits into two excitations,
respectively associated with the SW acoustic modes and the
emerging soft modes, gradually separating from each other and
getting thinner as the temperature goes down (see Fig.
\ref{fig:fig2}). The softening of the modes dies away to disappear
below $T/J=5\cdot 10^{-4}$. The soft mode, expected to be
non-dispersive in LSW theory, is here observed at finite energy,
due to the non linear nature of Eq. \ref{eq:eq_mvt}, which takes
account of the interactions between the SW. This effect is
expected to decrease with temperature, which is consistent with
the fact that the soft mode drops to zero energy when temperature
goes down (inset of Fig. \ref{fig:fig2}). Finally, one can discern
an additional peak at $\omega\simeq 2J$ for $T/J<2\cdot 10^{-3}$,
corresponding to optical modes.

Each mode $i$ of the excitation spectrum can be
characterized by its dispersion relation $\omega^i(\mathbf{Q})$,
its lifetime $\tau_{SW}^i\propto (\Gamma^i_{\mathrm{SW}})^{-1}$,
and its intensity $I_0^i$, all these quantities being accessible
by fitting the excitation spectrum at different temperatures and
$\mathbf{Q}_0$ values. Assuming Lorentzian shape for magnetic
excitations, the scattering function writes :
\begin{eqnarray}
S(\mathbf{Q}_0,\omega)
& = &
\sum_i \frac{I_0^i \Gamma^i}{(\Gamma^i)^2+(\omega \pm \omega^i(\mathbf{Q_0}))^2}
\end{eqnarray}
where $i$ runs over soft, acoustic and optical magnetic peaks for
a particular $\mathbf{Q}_0$ value.
\begin{figure}[t!]
\includegraphics[bb=10 10 191 238, scale=1.2]{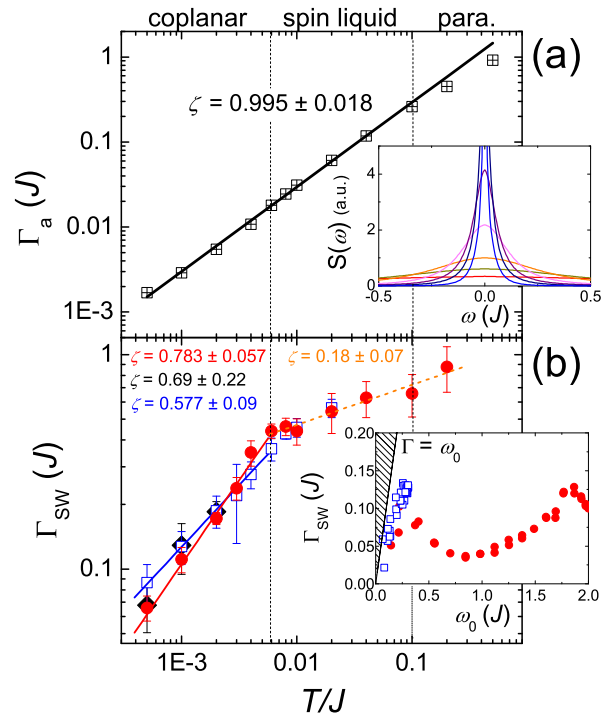}
\caption{(color online) (a) $\Gamma_a\propto\tau_a^{-1}$ obtained
by fitting the QE signal $S(\omega)$, shown in the inset at
several temperatures from $T/J=0.5$ to $5\cdot 10^{-4}$. (b)
$\Gamma_{\mathrm{SW}}\propto\tau^{-1}_{\mathrm{SW}}$ for the soft
(blue squares), in-plane (black diamonds) and out-of-plane (red
circles) acoustic modes for $\mathbf{Q}_0=2\pi(3/4,0)$. The error
bars have been obtained by averaging over different fitting
processes. If not seen, they are smaller than the symbols. The
inset displays $\Gamma_{SW}$ as a function
$\omega_0=\omega(\mathbf{Q})$, \textit{ie} for several
$\mathbf{Q}$-values. The hatched region forbids propagating
excitations.} \label{fig:fig3}
\end{figure}
We show Fig. \ref{fig:fig3}b the thermal variation of the
resulting SW HWHM $\Gamma_{\mathrm{SW}}\propto
\tau_{\mathrm{SW}}^{-1}$ for  $\mathbf{Q}_0 = 2\pi(3/4,0)$. It is
found out, contrarily to $\tau_a$ (see Fig. \ref{fig:fig3}a),
that $\tau_{\mathrm{SW}}$ follows two distinct regimes below and
above $T/J=5\cdot 10^{-3}$, both consistent with an algebraic law
$\tau_{\mathrm{SW}}= \mathcal{A}T^{-\zeta}$. $\tau_{\mathrm{SW}}$
is in principle reduced by two physical processes. The first,
common to all magnetic systems, is associated to the thermal
fluctuations and anharmonic interactions between SW modes. In a
disordered medium, this process is overwhelmed by a second one,
induced by the motion of the system between the different ground
states. In other words, even in the linear approximation, one is
left with a set of linear equations of motions with time dependent
initial conditions, the time variations of those being set by the
autocorrelation time $\tau_a$. In the pyrochlore antiferromagnet,
it has been shown that $\tau_a$ is also proportional to $T^{-1}$
and based on the above interpretation, it was proposed that the SW
lifetime $\tau_{\mathrm{SW}}$ is proportional to $T^{-1/2}$
\cite{moessner1998}. This behavior is expected at least in the
cooperative paramagnetic regime.
\begin{figure}[t!]
\includegraphics[scale=0.65]{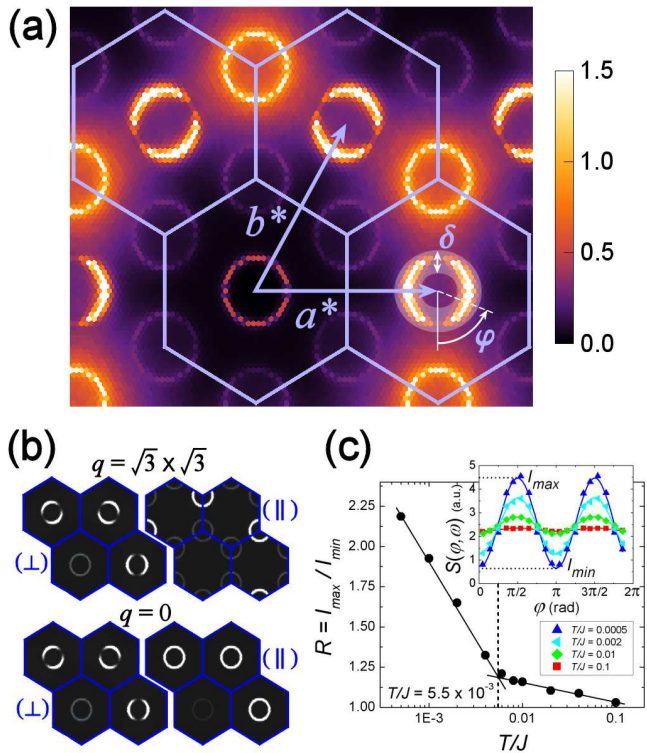}
\caption{(color online) (a) Intensity map $(a.u.)$ in reciprocal
space for $\omega=J$ and $T/J=5\cdot 10^{-4}$. The first and
neighboring BZ are in blue. (b) Out-of-plane ($\perp$)
\textit{(left)} and in-plane ($\parallel$) \textit{(right)}
components for \QSR\ and \QO\ spin configurations. (c) Anisotropy
parameter $R=I_{max}/I_{min}$ vs. temperature. Inset : scattering
function plotted for different temperatures in function of
$\varphi$, as defined in Fig. (a).} \label{fig:fig4}
\end{figure}
In the corresponding temperature range, we find $\zeta=0.18 \pm 0.07$, which is much lower than $1/2$. In the low $T$ regime, due to the selection of coplanar spin configurations, the out-of-plane $\omega^\perp (\mathbf{Q})$ and the in-plane $\omega^\parallel (\mathbf{Q})$ modes become different, which allows distinguishing the corresponding scattering process $S(\mathbf{Q},\omega)=S^\perp(\mathbf{Q} , \omega)+S^\parallel(\mathbf{Q},\omega)$. Within the LSW, or equivalently at very low temperatures, the out of plane scattering function $S^\perp(\mathbf{Q},\omega)$ is gauge invariant like, {\it i.e} does not depend on the three coloring state on top of
which the excitations develop (see Fig. \ref{fig:fig4}b, left). Conversely, the in-plane contribution $S^\parallel(\mathbf{Q},\omega)$ differs for each configuration it
is build on (see Fig. \ref{fig:fig4}b, right). In this coplanar regime, $\tau_{\mathrm{SW}}$ seems to behave in similar ways for the in-plane ($\zeta=0.69\pm 0.22$) and out-of-plane ($\zeta=0.783\pm 0.057$) acoustic modes, whereas $\zeta$ is slightly weaker for the soft mode ($\zeta=0.577\pm 0.09$).
Thereby, we clearly see two distinct dynamical regimes with a $\zeta$ value in the coplanar phase significantly larger than the value in the cooperative paramagnetic phase. This suggests that the SW lifetime is sensitive to the entropic selection of the coplanar manifold, that latter inducing some kind of stiffness in the spin texture. We can also wonder about the propagation of these magnetic excitations. If $\Gamma_{\mathrm{SW}}/\omega_0 < 1$ for a given mode, it can be considered as a propagative SW, its lifetime $\tau_{\mathrm{SW}}$ being longer than its period $\omega_0^{-1}$. This actually is the case for the soft and acoustic modes, at least up to $T/J \lesssim 0.1$ (see inset of Fig. 3b for $T/J = 5\cdot10^{-4}$). This underlines that SW propagation in the kagome antiferromagnet, although of different nature, is possible in both regimes of cooperative paramagnetism and entropy induced coplanarity.


We finally focus on the spectral weight distribution in
reciprocal space, which is non-uniform for the excitations
emerging from the BZ centers, {\it i.e} the out-of-plane acoustic
like modes. A two dimensional intensity map in reciprocal space
is shown in Fig. \ref{fig:fig4}a, for $\omega = J$ and $T/J = 5
\cdot 10^{-4}$. The spectral weight $S \left(\mathbf{Q},J\right)$
reaches its maximum value $I_{\max}$ along $\mathbf{a}^\star$,
$\mathbf{b}^\star$ or $\mathbf{a}^\star-\mathbf{b}^\star$ axis,
and fades out in each corresponding perpendicular direction (with
an intensity $I_{\min}$). This results in the presence of
``ghosts'' in the excitation rings, {\it i.e} existing excitations
with a strongly reduced cross section,   that would be invisible
{\it e.g} in neutron scattering experiments. Parameterizing these
rings by the angle $\varphi$ and integrating over a small width
$\delta$ of the ring (see Fig. \ref{fig:fig4}a) allows to
quantitatively analyze the spectral weight anisotropy as a
function of the temperature. Fig. \ref{fig:fig4}c displays the
evolution of the anisotropy parameter $R=I_{max}/I_{min}$ with the
temperature. Strong discrepancies between the cooperative
paramagnetic regime ($R \simeq 1$) and the entropy driven coplanar
regime, in which the anisotropy strongly increases with decreasing
temperature, are evidenced. At the lowest temperatures, where SW
propagates onto a disordered manifold, the strongly fluctuating spin texture could therefore be expected to drive these
extinctions. Actually, two distinct arguments rather
support a purely geometrical origin. Firstly, out of
plane excitations are gauge invariant like. Therefore, the
fluctuating nature of the manifold should not play any role.
Secondly, we have numerically computed $S \left( \mathbf{Q} ,
\omega \right)$ for configurations prepared in slightly distorted
ordered   \QO~ and \QSR~  phases and performed a LSW expansion
around these two phases (Fig. \ref{fig:fig4}b). All
calculations reproduce this spectral anisotropy, pointing out that
the ghost excitations rather originate from the peculiar geometry
of the lattice revealed by spin coplanarity.

In conclusion, the propagation of spatially structured collective
excitations has been numerically evidenced and quantitatively studied in the classical kagome
antiferromagnet. Although the SW exist
in both cooperative paramagnetic and coplanar regimes, their
lifetime was found very sensitive to the entropic
selection occurring below $T/J=5\cdot 10^{-3}$, in contrast with
the same inverse temperature dependence of the autocorrelation time in both regimes. At very low temperatures, these
propagative modes possess a noteworthy non uniform spectral
weight expressing effective dynamical extinction rules.

\acknowledgments We would like to thank C. Henley and E. Bonet for
helpful discussions. B.C. also thanks Shan-Ho Tsai for a useful
correspondence at the early stages of this work.



\begin{thebibliography}:

\bibitem{kagome} C. Zeng, and V. Elser, Phys. Rev B {\bf 42}, 8436 (1990);
C. Zeng, and V. Elser, Phys. Rev B {\bf 51}, 8318 (1995).

\bibitem{villain1979}J. Villain,
Z. Phys. B {\bf 33}, 31 (1979).

\bibitem{chalker1992}J. T. Chalker, P. C. W. Holdsworth, and E. F. Shender,
Phys. Rev. Lett. {\bf 68}, 855 (1992).

\bibitem{zhitomirsky2008}M. Zhitomirsky,
http://fr.arxiv.org/abs/0805.0676v1.

\bibitem{reimers1993}J. N. Reimers and A. J. Berlinsky,
Phys. Rev. B {\bf 48}, 9539 (1993).

\bibitem{huse1992}D. A. Huse and A. D. Rutenberg,
Phys. Rev. B {\bf 45}, 7536 (1992).

\bibitem{keren1994}A. Keren,
Phys. Rev. Lett. {\bf 72}, 3254 (1994)

\bibitem{tsai2000}S.-H. Tsai, A. Bunker, and D. P. Landau ,
Phys. Rev. B {\bf 61}, 333 (2000).

\bibitem{hukushima1996}K. Hukushima and K. Nemoto,
J. Phys. Soc. Jpn. {\bf 65}, 1604 (1996).

\bibitem{katzgraber2006}H. G Katzgraber, S. Trebst, D. A Huse and M. Troyer,
J. Stat. Mech.,  P03018 (2006).

\bibitem{harris1992}A. B. Harris, C. Kallin, and A. J. Berlinsky,
Phys. Rev. B {\bf 45}, 2899 (1992).

\bibitem{garanin1999}D. A. Garanin and B. Canals,
Phys. Rev. B {\bf 59}, 443 (1999).

\bibitem{moessner1998}R. Moessner and J. T. Chalker,
Phys. Rev. Lett. {\bf 80}, 2929 (1998); Phys. Rev. B {\bf 58}, 12049 (1998).

\end{thebibliography}
\end{document}